\documentclass[aps, prd, twocolumn, notitlepage, superscriptaddress]{revtex4-1}
\usepackage{graphicx}
\usepackage{float}
\usepackage{amsmath, amssymb}
\usepackage{color}
\usepackage{tensor}
\usepackage{epstopdf}
\usepackage{xcolor}
\usepackage{braket}	
\usepackage{hyperref}
\hypersetup{
    colorlinks,
    linkcolor={blue!50!black},
    citecolor={red!50!black},
    urlcolor={blue!80!black}
}

\usepackage{overpic}

\usepackage{soul}

\usepackage{multibib}
\newcommand{\be}{\begin{equation}}
\newcommand{\ee}{\end{equation}}
\newcommand{\ba}{\begin{eqnarray}}
\newcommand{\ea}{\end{eqnarray}}

\newcommand{\beq}{\begin{equation}}
\newcommand{\eeq}{\end{equation}}
\newcommand{\beqa}{\begin{eqnarray}}
\newcommand{\eeqa}{\end{eqnarray}}
\newcommand{\nn}{\nonumber}

\DeclareMathOperator{\tr}{tr} 			% for trace

\newcommand{\rh}{r_{\rm h}}
\newcommand{\kms}{T_{\rm KMS}}

\usepackage{colonequals} %Get a nice looking :=
\newcommand{\ce}{\colonequals}

\renewcommand{\Re}{\mathop{\rm Re}}		% for the real part
\begin{document}

\title{The BTZ black hole exhibits  anti-Hawking phenomena}

\author{Laura J. Henderson}
\email[]{l7henderson@uwaterloo.ca}
\affiliation{Department of Physics and Astronomy, University of Waterloo, Waterloo, Ontario, Canada, N2L 3G1}
\affiliation{Centre for Quantum Computation and Communication Technology,School of Science, RMIT University, Melbourne, Victoria 3001, Australia}

\author{Robie A. Hennigar}
\email[]{rhennigar@mun.ca}
\affiliation{Department of Mathematics and Statistics, Memorial University of Newfoundland, St. John’s, Newfoundland and Labrador, A1C 5S7, Canada}

\author{Robert B. Mann}
\email[]{rbmann@uwaterloo.ca}
\affiliation{Department of Physics and Astronomy, University of Waterloo, Waterloo, Ontario,  Canada, N2L 3G1}
\affiliation{Perimeter Institute for Theoretical Physics, 31 Caroline St. N., Waterloo, Ontario, Canada, N2L 2Y5}

\author{Alexander R. H. Smith}
\email[]{alexander.r.smith@dartmouth.edu}
\affiliation{Department of Physics and Astronomy, Dartmouth College, Hanover, New Hampshire 03755, USA}

\author{Jialin Zhang}
\email[]{jialinzhang@hunnu.edu.cn}
\affiliation{Department of Physics and Synergetic Innovation Center for Quantum Effects and Applications, Hunan Normal University, Changsha, Hunan 410081, China}

\date{\today}

\begin{abstract}
%The Unruh effect --- that accelerating observers perceive a thermal spectrum of particles --- is a cornerstone prediction of quantum field theory. 

 The Unruh effect is a surprising prediction of quantum field theory that asserts accelerating observers perceive a thermal spectrum of particles with a temperature proportional to their acceleration. However, it has recently been shown that particle detectors can click less often or even cool down as their acceleration increases, in contrast to the heating one would expect. This leads to the so called anti-Unruh phenomena. Here we consider detectors outside a BTZ black hole and demonstrate the existence of black hole analogues of these effects, which we dub anti-Hawking phenomena. 

%One of the most counter intuitive predictions of quantum field theory on curved spacetime is that black holes radiate and eventually evaporate. This radiation can in principle be verified in an operational manner by employing particle detectors that thermalize with this radiation. The flat space analogue of this situation is known as the Unruh effect: an observer uniformly accelerating through the Minkowski vacuum experiences a thermal bath of particles at a temperature that grows linearly with their acceleration. It has recently been shown that particle detectors can cool down as their acceleration increases, in contrast to the heating one would expect from Unruh effect. This leads to the so called anti-Unruh effect. By considering detectors outside the BTZ black hole, we demonstrate the black hole analogue of the effect which we dub the anti-Hawking effect.  \rbm{Sharpen this}
\end{abstract}


\maketitle

The Unruh effect is one of the most striking predictions of quantum field theory. Observers undergoing uniform acceleration in the Minkowski vacuum will experience a thermal bath of particles at a temperature proportional to the acceleration~\cite{Unruh:1976}.  The Unruh effect is the flat spacetime `cousin' of the Hawking effect, which describes  black hole evaporation. In both instances,  thermality of the Unruh/Hawking radiation can be verified via local measurements described by  particle detector models that interact with the field, thermalizing in the limit of infinite time~\cite{Birrell:1982ix,Hodgkinson:2012}.

Recently it has been realized that probing the Unruh effect via detector models can lead to counter-intuitive results that become manifest under quite generic circumstances~\cite{Brenna:2015fga,Garay:2016cpf}.  It has been shown that a detector can click \textit{less often} as the temperature of the field increases, and this can persist even in the limit of infinite interaction time. For finite interaction times that are long enough that the detector can still be regarded as having approximately thermalized, it is possible for the temperature recorded by the detector to \textit{decrease} as the temperature of the field \textit{increases}. Collectively these results have been termed {\it anti-Unruh phenomena}, with the former corresponding to the \textit{weak anti-Unruh effect} and the latter the \textit{strong anti-Unruh effect}~\cite{Brenna:2015fga,Garay:2016cpf}.

Unfortunately, relatively little is known about the general conditions that give rise to these effects and even less is known in scenarios with non-trivial spacetime curvature. Most significantly, anti-Unruh phenomena are inherent to accelerating motion and are not observed by an inertial detector coupled to a thermal state ~\cite{Garay:2016cpf}.  The effects are present for topological qubits undergoing various motions in Minkowski space and have been found to be associated with a decoherence impedance effect~\cite{Liu:2016ihf}.  It has also been shown~\cite{Li:2018xil} that tuning to parameter regimes where anti-Unruh phenomena are active can serve as a mechanism to amplify the amount of entanglement extracted from the quantum vacuum in entanglement harvesting protocols~\cite{Valentini:1991, Reznik:2002fz, Reznik:2005, Pozas-Kerstjens:2015}.

% It was shown in~\cite{Liu:2016ihf} that the effect is present for couplings to fermionic fields and different types of detector motion. Moreover it was found in~\cite{Liu:2016ihf} that anti-Unruh phenomena is associated with a \textit{decoherence impedance} 

Based on the similarity between the Unruh and Hawking effects, it is natural to inquire if there are analogues of anti-Unruh phenomena for the Hawking effect.  We consider this question here
 and find that  detectors  outside a black hole can experience an analogous  \emph{anti-Hawking effect}. Specifically, we consider the case of the ($2+1$)-dimensional BTZ black hole~\cite{Banados:1992,Banados:1993}, and analyze the response of a detector interacting with a (massless) conformally coupled  scalar field in this background. To ensure that the effects we observe are truly due to the black hole (and are not simply an effect of an acceleration horizon), we will also analyze a physically comparable set-up involving an accelerating detector in AdS$_3$.

As a simplified model of an atom interacting with the vacuum, we employ the Unruh-DeWitt detector \cite{Unruh:1976, DeWitt:1979}, which consists of a two-level quantum system moving along the spacetime trajectory $x_D(\tau)$, parametrized by the detector's proper time $\tau$, that interacts locally with a scalar field $\phi(x)$.  The ground and excited states of the detector are denoted as $\ket{0_D}$ and $\ket{1_D}$, respectively, and separated by an energy gap $\Omega$. In the interaction picture, the Hamiltonian describing the interaction of the detector with the field is
\begin{align}
H_D(\tau) \!&=\! \lambda \chi_D\! \left(\tau \right)\!\Big(e^{ i\Omega \tau}  \sigma^+  +  e^{- i\Omega \tau} \! \sigma^- \Big) \otimes  \phi\left[x_D(\tau)\right], \label{InteractionHamiltonian}
\end{align}
where $\chi_D(\tau) \leq 1 $ is a switching function controlling the duration of the interaction, and $\sigma^+ \ce  \ket{1_D}\!\bra{0_D}$ and $\sigma^- \ce  \ket{0_D}\!\bra{1_D}$ are ladder operators acting on the Hilbert space associated with the detector. Although simple, this model captures the relevant features of the light-matter interaction when no angular momentum exchange is involved \cite{Martin-Martinez2013, Alvaro}.

%Consider a detector moving with with trajectory $x_D(\tau)$ parametrized by the detector's proper time $\tau$.  
Suppose the detector is initially ($\tau \to -\infty)$ in its ground state while the field is initially in the its vacuum state $\ket{0}$, so that the joint state of the detector and field together is $\ket{\Psi_i} = \ket{0_D} \ket{0}$. Given the field/detector interaction Hamiltonian \eqref{InteractionHamiltonian}, the final ($\tau \to \infty$) state of the field-detector system is given by
\begin{align}
\ket{\Psi_f} = \mathcal{T} e^{  -i \int  dt\, \left[ \frac{d \tau}{dt} H_D(\tau) \right] } \ket{\Psi_i}, \label{totalfinalstate}
\end{align}
where $\mathcal{T}$ is the time ordering operator and we have chosen to evolve the field and detectors with respect to an appropriate coordinate time $t$ with respect to which the vacuum state of the field is defined. The final state of the detector alone is obtained from~\eqref{totalfinalstate} by tracing out the field degrees of freedom, $\rho_{A} \ce \tr_\phi \big( \ket{\Psi_f}\!\bra{\Psi_f} \big)$, which to leading order in the interaction strength is given by
\begin{align}
\rho_D &\ce\begin{pmatrix}
1-P_D & 0 \\
0 & P_D
\end{pmatrix}
+\mathcal{O}\!\left( \lambda^4\right)  ,
\label{reducedD}
\end{align}
where 
\begin{align}
P_D \ce \lambda^2 \int d\tau  &d \tau' \, \chi_D(\tau) \chi_D(\tau') 
\nn\\
&\times e^{-i \Omega \left(\tau-\tau'\right)} W\!\left(x_D(t) ,x_D(t')\right), \label{PJ}  
\end{align}
where $W(x,x')\ce\bra{0} \phi(x) \phi(x') \ket{0}$ is the Wightman function associated with the field state $\ket{0}$. The object $P_D$ is the \textit{transition probability}, though in this work it will be more convenient to work with the \textit{response function} 
\be 
{\cal F} \ce \frac{P_D}{\lambda^2 \sigma} \, ,
\ee
where $\sigma$ is a characteristic time scale for the interaction --- we shall provide  an explicit definition for $\sigma$ 
below, when we introduce a specific switching function.

We focus on quantum field theory states that satisfy the Kubo-Martin-Schwinger (KMS) condition~\cite{Kubo:1957mj, Martin:1959jp, Haag:1967sg}. The KMS condition provides a general definition of thermal states in quantum field theory where the usual Gibbs distribution may be problematic or difficult to rigorously define. For a KMS state  with temperature $T_{\rm KMS}$, the corresponding Wightman function will satisfy the following imaginary time boundary condition~\footnote{{When the Wightman function is \textit{stationary}, i.e. $W(\tau, \tau') = W(\Delta \tau)$,  then the KMS condition reduces to `anti-periodicity' in imaginary time: $W(\Delta \tau - i \beta) = W(-\Delta \tau)$.}} 
$
W(\tau - i/T_{\rm KMS}, \tau') = W(\tau', \tau) 
$,
where we have introduced the shorthand $W(\tau', \tau) \ce W(x_D(\tau'), x_D(\tau))$.

To understand the  process by which the detector thermalizes with the field it is useful to introduce the excitation to de-excitation ratio (EDR) of the detector:
\be 
{\cal R} = \frac{{\cal F}(\Omega)}{{\cal F}(-\Omega)}  .
\ee
If the detector thermalizes to a temperature $T$, then this ratio will satisfy the detailed balance form of the KMS condition~\cite{Fewster:2016ewy} 
\be 
{\cal R} = e^{-\Omega/ T},
\ee
with the temperature independent of the gap. We can define a temperature estimator  from the EDR ratio that we will denote as $T_{\rm EDR}$:
\be 
T_{\rm EDR} = - \frac{\Omega}{\log {\cal R}}  .
\ee

The technical considerations below will require only the details of a conformally coupled scalar on AdS$_3$.  For convenience, we collect the relevant details in the appendix, though a more detailed discussion using the same notation can be found in~\cite{Henderson:2017yuv, Henderson:2018lcy}.

Let us begin by considering accelerating observers in AdS$_3$. Contrary to the situation in flat space, an accelerating observer in AdS need not see an acceleration horizon. There exists a critical acceleration $a_c = 1/\ell$ partitioning observers into three classes: sub-critical ($a < a_c$), critical $(a = a_c)$, and super-critical ($a > a_c$). It is only the latter that experiences an acceleration horizon, and it this class of super-critical accelerations that we shall consider here~\footnote{See Ref.~\cite{Jennings:2010vk} for a discussion of the response rate of detectors in the critical and sub-critical cases.}. In this case the metric reads
\be\label{AdSRin}
ds^2 = -\left(\frac{r^2}{\ell^2} - 1 \right) dt^2 + \left(\frac{r^2}{\ell^2} - 1 \right)^{-1} dr^2 + r^2 d\Phi^2 \, . 
\ee
This is the AdS-Rindler metric,  which contains an acceleration horizon located at $r = \ell$. Here we should note the coordinate $\Phi$ takes on values on the full real line.

%s we will see below, this metric is essentially the BTZ metric with mass parameter $ M = 1$, but there is an important difference. Here, the coordinate $\Phi$ \textit{is not} identified with a period of $2\pi$. Here $\Phi$ is a non-compact coordinate with range $\Phi \in (-\infty, \infty)$. 

An observer at constant $r=R_D$ has an acceleration with magnitude given by 
\be 
\label{eqn:RindAcc}
|a| = \frac{1}{\ell} \frac{x}{\sqrt{x^2-1}} \, , \quad x := \frac{R_D}{\ell} .
\ee
The minimum acceleration is $|a| = 1/\ell$ (the critical acceleration $a_c$), and this happens near $r = \infty$, while $|a| \to \infty$ as $r \to \ell$. 

%We will choose the switching functions of the detectors to be Gaussian, $\chi_D(\tau) = \exp\left(-\tau^2/2\sigma^2\right)$,

We choose for the detector a Gaussian switching function $\chi_D(\tau) = \exp\left(-\tau^2/2\sigma^2\right)$, with the interpretation that the detector interacts with the field for an interval of time $\sim \sigma$ centered on the $t=0$ hypersurface. To determine the KMS temperature of the field, it is easy to show that regularity of the Euclidean sector requires the imaginary time $t$ has period $\beta = 2 \pi \ell$, from which it follows that the temperature is $T = 1/\beta = 1/(2 \pi \ell)$.  However, to compute the local temperature of the field at the location of the detector we must also account for time dilation effects. Doing so, we obtain $T_{\rm KMS} = \sqrt{a^2 \ell^2 - 1}/(2\pi \ell )$,
where we have used Eq.~\eqref{eqn:RindAcc} to write the temperature in terms of $a$ instead of $R_D$. In terms of these variables, we find (see the appendix) the response function of a   static detector at fixed $r$ and $\Phi$ is given by 
\begin{widetext}
\begin{align}
{\cal F}_{\rm AdS-R} =  \frac{\sqrt{\pi}}{4} - \frac{i}{4 \sqrt{\pi}} {\rm PV} \int_{-\infty}^\infty dz \frac{e^{-z^2/(2 \pi T_{\rm KMS}\sigma)^2} e^{- i \Omega  z/ (\pi T_{\rm KMS})}}{\sinh z} - \frac{\zeta}{2 \sqrt{2 \pi}} {\rm Re}  \int_{0}^\infty dz \frac{e^{- z^2/(4 \pi T_{\rm KMS}\sigma)^2} e^{-i \Omega z/ (2 \pi T_{\rm KMS})}}{\sqrt{1 + 8 \pi^2 \ell^2 T_{\rm KMS}^2 - \cosh z }} ,
\label{RindResponse}
\end{align}
\end{widetext}
where {\rm PV} means that the principle value of the integral should be taken and $\zeta$ specifies the boundary condition satisfied by the field at spatial infinity: $\zeta = -1$ (Neumann),  $\zeta = 0$ (transparent), and  $\zeta = 1$ (Dirichlet).
%\be 
%{\cal F} = \frac{\sqrt{\pi}}{4} -  \frac{i}{4 \sqrt{\pi}} {\rm PV} \int_{-\infty}^\infty dz \frac{\exp \left[\frac{-\ell^2 z^2}{\sigma^2 \left(a^2 \ell^2-1\right)} \right] \exp \left[- \frac{ 2 i \Omega \ell z }{\sqrt{a^2\ell^2-1}} \right]}{\sinh z} - \frac{\zeta}{2 \sqrt{2 \pi}} {\rm Re} \int_{0}^\infty dz \frac{\exp\left[-\frac{ \ell^2 z^2}{4 \sigma^2 (a^2\ell^2 - 1)}  \right] \exp\left[ - \frac{i \Omega \ell z}{\sqrt{a^2 \ell^2 - 1}}  \right]}{\sqrt{2a^2\ell^2 - 1 - \cosh z}}  \, .
%\label{RindResponse}
%\ee

%Note that this is identical to the $n=0$ term of the transition probability for the BTZ black hole with $M=1$. This is not surprising --- it is the identifications that distinguish the BTZ black hole from an acceleration horizon, and the $n=0$ term in the BTZ image sum simply is the acceleration horizon contribution. 

We can confirm that in the limit of infinite interaction time the detector truly thermalizes to $T_{\rm KMS}$. To see this, note that in the limit $\sigma \to \infty$ the integrals appearing in Eq.~\eqref{RindResponse}  can be written in terms of special functions:
\begin{align}
\label{rindlerResponseA} 
{\cal F}^{\sigma \to \infty}_{\rm AdS-R} &= \frac{\sqrt{\pi}}{4} \left[1 - \tanh \left(\frac{\Omega}{2 \kms} \right) \right]
\nn\\
&\times\left\{1   - \zeta  P_{-\frac{1}{2} + \frac{i \Omega}{2 \pi \kms}} \left(1 + 8 \pi^2 \ell^2 \kms^2 \right)  \right\} ,
\end{align}
where $P_{\nu}$ is the associated Legendre function of the first kind and satisfies $P_{-1/2 + i \lambda} = P_{-1/2 - i \lambda}$. It then follows that for all boundary conditions the detector satisfies the detailed balance condition with \mbox{$T = T_{\rm KMS}$}.

\begin{figure}%[!ht]
\begin{overpic}[width=0.4\textwidth]{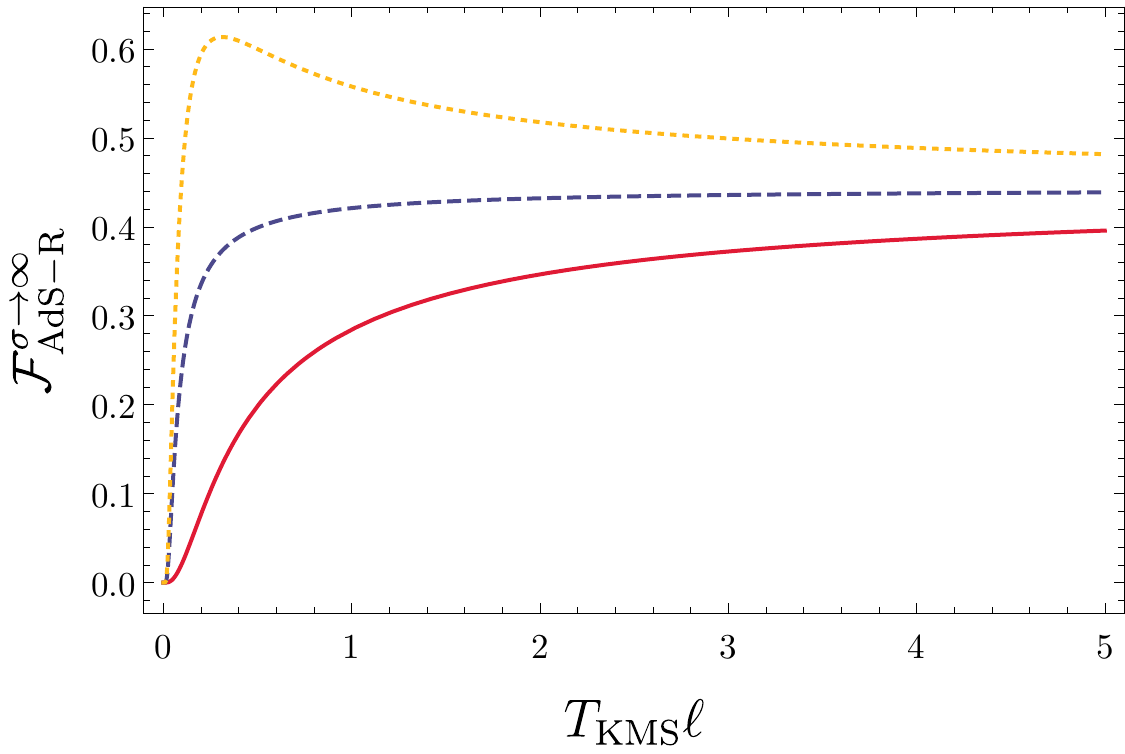}\put(55,15){
\includegraphics[scale=0.5]{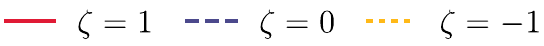}}
\end{overpic}
\caption{Weak anti-Unruh effect for AdS-Rindler. Here we show a plot of the response function against the KMS temperature of the field for the particular choice of $\Omega \ell = 1/10$ in the infinite interaction limit. The weak anti-Unruh effect is present only for Neumann boundary conditions.}
\label{fig:WAU-AdSR}
\end{figure}

From Eq.~\eqref{rindlerResponseA} we can extract some details concerning weak anti-Unruh phenomena, as indicated Fig.~\ref{fig:WAU-AdSR}, which plots all three boundary conditions.
For the $\zeta=-1$ Neumann boundary condition we see a region of $T_{\rm KMS}$ values for which the response function \textit{decreases} with increasing $T_{\rm KMS}$. This indicates that  weak anti-Unruh phenomena are present in the infinite interaction limit. A more detailed numerical exploration of the parameter space strongly suggests that it is \textit{only} in the case of Neumann boundary conditions that this can occur, irrespective of the energy gap \footnote{If we allow for arbitrary $\zeta \in [-1, 1]$, then we find strong evidence that $\zeta < 0$ is required to observe this weak anti-Unruh effect in the infinite interaction time limit, with larger gap restricting the allowed parameter space to more negative $\zeta$.}. Increasing the detector gap has the effect of pushing the region where $\partial {\cal F}/\partial T_{\rm KMS} < 0$ to higher $T_{\rm KMS}$ --- or, in other words, a detector with a larger gap needs to be closer to the AdS-Rindler horizon to observe the effect. We find that the effect is absent in flat spacetime, where these boundary terms vanish, consistent with~\cite{Garay:2016cpf}, where it was claimed that both strong and weak anti-Unruh phenomena are absent for accelerating observers coupled to a massless field in $(2+1)$-dimensional Minkowski space.  

We have also explored numerically the behaviour of the response function for finite interaction times. 
Provided $\Omega \ell^2 < \sigma$ and the boundary conditions are Neumann, we find that the weak anti-Unruh phenomenon emerges; we find no examples of weak anti-Unruh phenomena for other boundary conditions. Furthermore, despite an intensive exploration of the parameter space, we  find no examples of \textit{strong} anti-Unruh phenomena, though since our methods are numerical we cannot completely rule out this possibility.

Turning to black holes, we consider  the BTZ  black hole of mass $M$, whose metric is
\be 
ds^2 = - f(r) dt^2 + \frac{dr^2}{f(r)} + r^2 d\phi^2 ,
\ee
where $f(r) = r^2/\ell^2 - M$ and  the horizon is located at $\rh = \sqrt{M} \ell$. In these coordinates, the angular direction is identified $\phi \sim \phi + 2 \pi$, and it is ultimately this identification that distinguishes the BTZ metric from the AdS-Rindler case \eqref{AdSRin}. Due to the simplicity of the BTZ spacetime, the Wightman function for a conformally coupled scalar field (in the Hartle-Hawking state) in a BTZ background is known analytically~\cite{Lifschytz:1994,Carlip:2003}. We provide the necessary details of this construction in the appendix.

%Armed with these details, we can obtain an expression for the response function of a detector located at constant $r$ and $\phi$ in the following way~\cite{Lifschytz:1994, Hodfkinson:2012,Hodfkinson:2012a,Smith:2014}, which is done in Appendix~\ref{BTZAppendix}. The key observation is that while the $n=0$ terms in the BTZ response function coincide with the AdS-Rindler result, the other terms appearing in the image sum do not, and it is these terms that account for `bonefide black hole' effects. These higher-order terms are controlled by the BTZ mass parameter $M$, and their effect is most pronounced when $M$ is small.

We can obtain an expression for the response function of a detector located at constant $r$ and $\phi$ following~\cite{Lifschytz:1994, Hodfkinson:2012,Hodfkinson:2012a,Smith:2014}.  This is done in Appendix~\ref{BTZAppendix} and we mention only a few relevant details here. To facilitate comparison with the AdS-Rindler case, we note that the acceleration of a constant $r$ trajectory in the BTZ spacetime is
\be\label{BTZacc} 
|a| = \frac{1}{\ell} \frac{y}{\sqrt{y^2-1}} \, , \quad y := \frac{R_D}{\ell \sqrt{M}} \, .
\ee
which is exactly the same as in the AdS-Rindler case expressed in \eqref{eqn:RindAcc}, modulo a rescaling of the parameter $x = R_D/\ell$. The Hawking temperature can be computed using the usual Euclidean trick: $T_{\rm H} =  \sqrt{M}/(2 \pi \ell)$. Taking into account time dilation effects, the local KMS temperature calculated at the position of detector on an $r = R_D$ surface is  given by $T_{\rm KMS} = \sqrt{a^2 \ell^2 - 1}/(2 \pi \ell)$, identical to the AdS-Rindler case,
where we have used \eqref{BTZacc}. The response function is then:
\begin{widetext}
\begin{align}
{\cal F}_{\rm BTZ} = {\cal F}_{\rm AdS-R} +&\frac{1}{\sqrt{2 \pi}} \sum_{n=1}^{\infty} \Bigg\{\int_{0}^{\infty} dz\ {\rm{Re}}\Bigg[ \frac{\exp\big(-z^2/(4 \pi \sigma T_{\rm KMS})^2\big)\exp\big(-i \Omega z/(2 \pi T_{\rm KMS}) \big)}{\sqrt{\cosh\big(\alpha_{n}^-\big)-\cosh(z)}}\Bigg] 
\nn\\
&- \zeta\int_{0}^{\infty} dz\ {\rm{Re}}\Bigg[\frac{\exp\big(-z^2/(4 \pi \sigma T_{\rm KMS})^2\big)\exp\big(-i\Omega z/(2 \pi T_{\rm KMS})\big)}{\sqrt{\cosh\big(\alpha_{n}^+\big)-\cosh(z)}}\Bigg]\Bigg\} \, ,
\end{align}
where 
\begin{align}
\cosh \alpha^\mp_n &=  \left(1 + 4 \pi^2 \ell^2 T_{\rm KMS}^2 \right)\cosh (2 \pi n \sqrt{M})  \mp 4 \pi^2 \ell^2 T_{\rm KMS}^2 \, .
\end{align}
\end{widetext}
The key observation is that, while the $n=0$ terms in the BTZ response function coincide exactly with the AdS-Rindler result, the remaining terms are novel. It is then these terms that are responsible for bonefide black hole effects. These higher-order terms are controlled by the BTZ mass parameter $M$, and their effect is most pronounced when $M$ is small.

\begin{figure}%[!ht]
\begin{overpic}[width=0.4\textwidth]{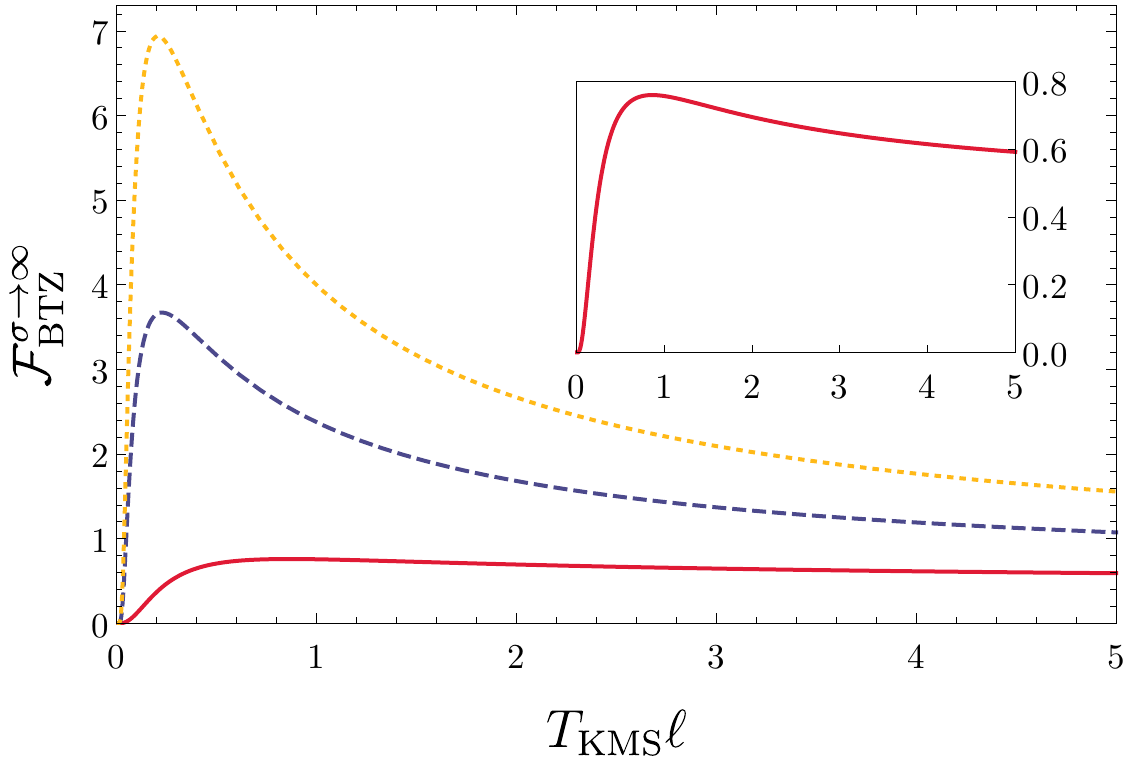}\put(18,62){
\includegraphics[scale=0.5]{Legend.pdf}}
\end{overpic}
\caption{Here we show a plot of the BTZ response function in the infinite interaction time limit against the KMS temperature of the field. In this plot, $M = 1/100$ and $\Omega \ell = 1/10$. The inset shows a zoomed version of the Dirichlet boundary condition curve. For this choice of parameters, the weak anti-Unruh effect is observed for all three boundary conditions.}
\label{fig:BTZ-WAU}
\end{figure}

Once again, in the limit of infinite interaction time, the BTZ response function can be expressed explicitly in terms of special functions as
\begin{align} 
{\cal F}_{\rm BTZ}^{\sigma \to \infty} &= \frac{\sqrt{\pi}}{4} \left[1 - \tanh\left(\frac{\Omega}{2 T_{\rm KMS}}\right) \right] 
\nn\\
&\quad \times\sum_{n=-\infty}^{n = \infty} \left[P_{-\frac{1}{2} + \frac{i \Omega}{2 \pi T_{\rm KMS}}} \left( \cosh \alpha_n^- \right) \right.
\nn\\
&\quad \left.- \zeta  P_{-\frac{1}{2} + \frac{i \Omega}{2 \pi T_{\rm KMS}}} \left( \cosh \alpha_n^+ \right) \right], 
\end{align}
From this expression, and using the fact that (for real $\lambda$) $P_{-1/2 + i \lambda} = P_{-1/2 - i \lambda}$, it is straightforward to show that the detector satisfies the detailed balance condition with $T = T_{\rm KMS}$. Plotting the response function in Fig.~\ref{fig:BTZ-WAU}, we see that 
weak anti-Unruh phenomena are present provided  $M$ is small enough. This holds for
 all three boundary conditions in strong contrast to the 
AdS-Rindler case.

 \begin{figure}[!ht]
 \begin{overpic}[width=0.4\textwidth]{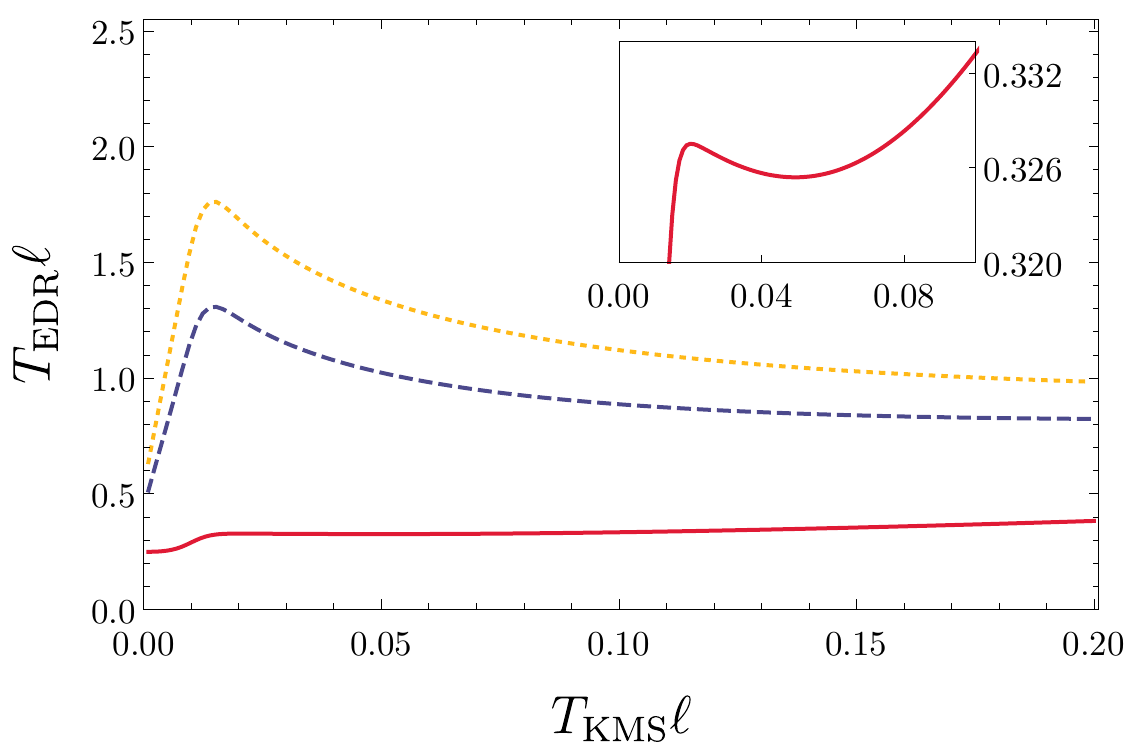}\put(13,58){
\includegraphics[scale=0.5]{Legend.pdf}}
\end{overpic}
\ \\ \ \\
\begin{overpic}[width=0.4\textwidth]{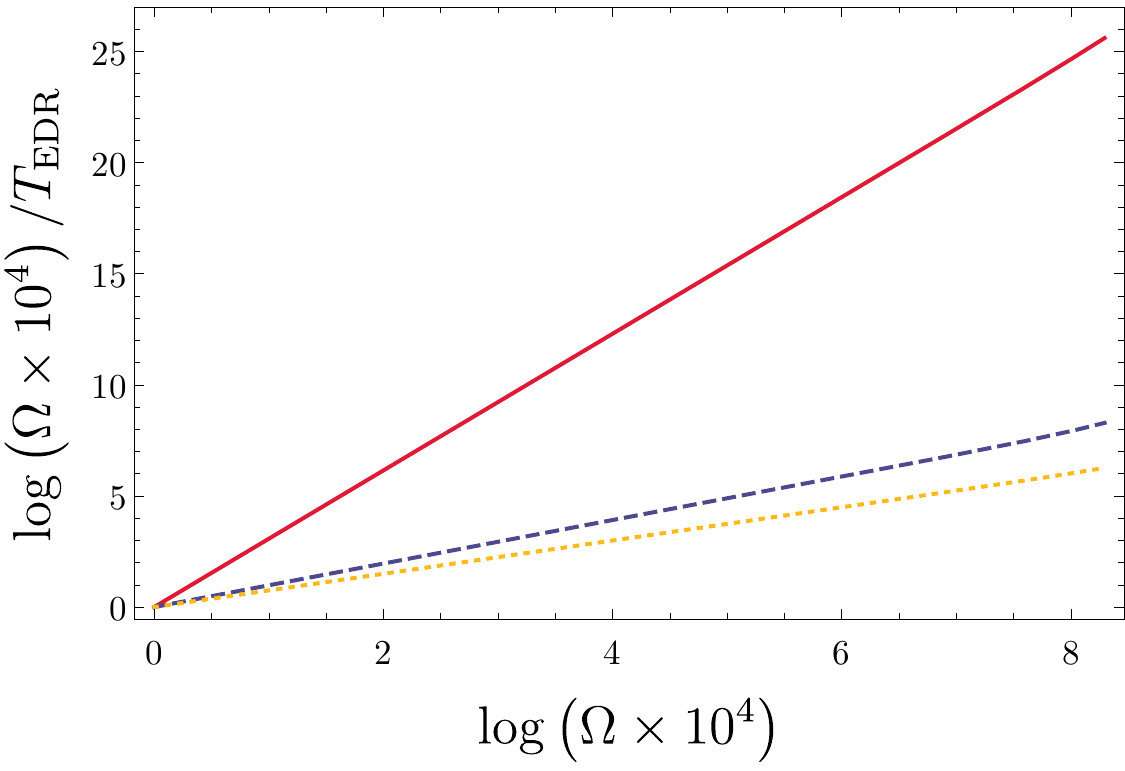}\put(13,60){
\includegraphics[scale=0.5]{Legend.pdf}}
\end{overpic}
\caption{{\it Top}: Strong `anti-Hawking' effect for BTZ black hole. Here $M = 1/500$, $\Omega \ell  = 1/100$, $\sigma/\ell = 1$. The inset shows the $\zeta=1$ case in greater detail, revealing the strong version of the effect. {\it Bottom}: Gap independence of the EDR temperature in BTZ spacetime. In this plot we have set $M = 1/500$, $T_{\rm KMS}  = 0.05$, $\sigma/\ell = 1$.}
\label{fig:StrongBTZ}
\end{figure}

Even more remarkably we find at  finite interaction times the appearance of a \textit{strong} version of  the phenomenon, in which an {\it increasing} KMS temperature of the field yields {\it decreasing} EDR temperature.
In terms of the physical spacetime, this corresponds to a detector registering a lower EDR temperature as it gets closer to the black hole, as shown in Fig.~\ref{fig:StrongBTZ}.  We find that the strong effect `emerges' for small values of  $M$. Though it is not shown in the figure, we find that $T_{\rm EDR}$ always grows with $T_{\rm KMS}$ for large values of $T_{\rm KMS}$.

Unfortunately, since the detector is at once both accelerating and in a black hole spacetime, we cannot strictly speaking separate the Unruh and Hawking effects experienced by the detector. However, we see that in the same regions of parameter space this effect is absent for a static detector in AdS$_3$, while present for a  static detector outside  a BTZ black hole. We emphasize that this indicates that the strong version of the effect shown here \textit{is} a consequence of the black hole, and is not simply due to the acceleration horizon. In this sense, it seems appropriate to refer to it as an `anti-Hawking effect'.
Furthermore, the effect is present for all boundary conditions on the field, though  most pronounced for the Neumann case.

Since the interaction time is finite, the detector no longer satisfies the detailed balance condition for all energy gaps $\Omega$. However, we have confirmed that in the regions of parameter space where the effect is observed, the EDR temperature is effectively gap-independent for a wide range of values of $\Omega$, indicating that the detector can be considered to have thermalized in an approximate sense~\cite{Fewster:2016ewy, Garay:2016cpf} --- this is shown explicitly in Fig.~\ref{fig:StrongBTZ} for $\Omega$'s ranging over four orders of magnitude. Of course, for a given choice of (finite) interaction time, the EDR temperature \textit{will} exhibit energy gap dependence for sufficiently large values of $\Omega$.

Summarizing, we have studied the response function of detectors in three-dimensional AdS-Rindler space and the BTZ spacetime for interactions of both finite and infinite duration. In both cases, the Wightman function of the field pulled back to the trajectory of the detector satisfies the KMS condition, and moreover in the limit of infinite interaction time the detectors thermalize to the KMS temperature of the field.  

For  AdS-Rindler, we have found that in this limit  \textit{weak} anti-Unruh phenomena is present, provided that the field satisfies Neumann boundary conditions. This counter-intuitive effect is characterized by the detector registering fewer clicks as the temperature of the field increases.  Our result indicates that spacetime curvature can induce this effect in cases where it would be absent in flat spacetime~\cite{Garay:2016cpf}. 

For the BTZ case, we find that in the limit of infinite interaction time the same effect is present for all choices of boundary conditions, provided that the black hole mass $M$ is sufficiently small. This provides evidence of a weak anti-Hawking effect.  We also found  for the first time evidence of a strong anti-Hawking effect for finite detector interaction times, provided that its mass $M$ is sufficiently small. This effect, present for all boundary conditions, corresponds to the EDR temperature registered by the detector \textit{decreasing} as the KMS temperature of the field \textit{increases}. This is the black hole analog of the strong anti-Unruh effect discussed in~\cite{Brenna:2015fga, Garay:2016cpf}. Just as in those cases, due to the finite nature of the interaction, the EDR temperature is not completely independent of the detector's energy gap. Nonetheless, we demonstrated that over the range of parameters for which the effect is observed, the EDR temperature is effectively independent of the energy gap, provided that it is sufficiently small.

There are a few natural directions that could be pursued in light of our results. The most obvious --- and perhaps also most interesting --- is to study higher dimensional black holes for evidence of these effects. In the same spirit, it would be interesting to verify whether or not the effects persist for the case of infalling rather than static detectors. In this circumstance, observers are necessarily restricted to finite interaction times since their world-lines will ultimately terminate at the singularity.  It would also be worthwhile to determine the implications of the anti-Hawking phenomena in the context of entanglement harvesting and decoherence, as for example in~\cite{Liu:2016ihf,Li:2018xil, Henderson:2017yuv}. Additionally, since these effects are absent for inertial detectors interacting with thermal states~\cite{Garay:2016cpf}, it is natural to wonder if they can occur in the context of the Gibbons-Hawking effect~\cite{Gibbons:1977mu}. Finally, it would seem to be important to determine exactly what physical characteristics of the underlying fields are necessary or sufficient for a detector to observe these effects.

{\bf Acknowledgments} --- We would like to thank Eduardo Mart{\'i}n-Mart{\'i}nez and Jos{\'e} de Ram{\'o}n for useful discussions and feedback on an earlier draft of this manuscript. This work was supported in part by the Natural Sciences and Engineering Research Council of Canada (NSERC) and the Dartmouth College Society of Fellows. R. A. Hennigar acknowledges the support of the NSERC Banting Postdoctoral Fellowship programme. The author J. Zhang thanks for the support from the National Natural Science Foundation of China under Grants No.11435006 and No.11690034.
\bibliography{HawkingBib}

\onecolumngrid  \vspace{1cm}
\newpage
\begin{center}
{\Large\bf Appendices}
\end{center}
\appendix
\tableofcontents

\section{Geometric and field theoretic details}

In this appendix we collect a number of useful but well-established results concerning the AdS$_3$ and BTZ geometries, and expand upon some presently  rather briefly in the main text.

The AdS$_3$ geometry with cosmological length scale $\ell$ can be obtained as the induced metric on a $3$-dimensional hyperboloid
\begin{align}
X_1^2+X_2^2-T_1^2-T_2^2=-\ell^2,
\label{hyp1}
\end{align}
embedded in a flat 4-dimensional geometry
\begin{align}\label{ds4}
dS^2=  dX_1^2 + dX_2^2  -dT_1^2 - dT_2^2,
\end{align}
with coordinates $(X_1, X_2, T_1, T_2)$~\cite{Carlip:2003}. This geometry solves the $(2+1)$-dimensional Einstein equations with negative cosmological constant $\Lambda = -1/\ell^{2}$.  The AdS-Rindler and BTZ spacetimes in the coordinate charts used in the main text can be obtained directly from this embedding function with suitible choices for $T_1, T_2, X_1, X_2$ and ---in the BTZ case--- suitable identifications.

In the case of AdS-Rindler, the transformation
\begin{eqnarray}
T_1 &=& \ell \sqrt{\frac{r^2}{\ell^2}} \cosh \Phi \, ,
    \quad
X_1 =   \ell \sqrt{\frac{r^2}{\ell^2}} \sinh \Phi \, ,
    \nn\\
T_2 &=& \ell \sqrt{\frac{r^2}{\ell^2}-1} \sinh \frac{t}{\ell} \, ,
    \quad
X_2 = \ell \sqrt{\frac{r^2}{\ell^2}-1} \cosh \frac{t}{\ell} \, .
\end{eqnarray}
produces the metric presented in the main text, while the transformation
\begin{align}
T_1 &= \ell \sqrt{\frac{r^2}{M \ell^2}} \cosh (\sqrt{M} \phi) \, , \quad
X_1 =   \ell \sqrt{\frac{r^2}{M \ell^2}} \sinh (\sqrt{M} \phi) \, ,
\nn\\
T_2 &= \ell \sqrt{\frac{r^2}{M \ell^2}-1} \sinh \frac{\sqrt{M} t}{\ell} \, ,    \quad
X_2 = \ell \sqrt{\frac{r^2}{M \ell^2}-1} \cosh \frac{\sqrt{M} t}{\ell} \, .
\end{align}
followed by the identification $\phi \sim \phi + 2 \pi$ produces the BTZ metric presented in the main text.

For a (massless) conformally coupled scalar field living on the AdS$_3$ geometry, the vacuum Wightman function is~\cite{Carlip:2003}
\begin{align}
W_{\rm AdS}^{(\zeta)}(x,x')=\frac{1}{4\pi\ell\sqrt{2}} \left(\frac{1}{\sqrt{\sigma(x,x')}}-\frac{\zeta}{\sqrt{\sigma(x,x')+2}}\right), \label{wightmanf}
\end{align}
where
\begin{align}
\sigma(x,x') &= \frac{1}{2\ell^2} \left[ \left(X_1-X_1' \right)^2- \left(T_1-T_1'\right)^2 +\left(X_2-X_2'\right)^2 - \left(T_2-T_2'\right)^2 \right], \label{deltasigma}
\end{align}
is the square distance between $x$ and $x'$ in the embedding space $\mathbb{R}^{2,2}$.
The parameter $\zeta \in \{1,0,-1\}$ respectively specifies  Dirichlet ($\zeta = 1$), transparent ($\zeta = 0$), and Neumann ($\zeta = -1$) boundary conditions satisfied by the field at spatial infinity. Furthermore,  the Hartle-Hawking vacuum on the BTZ black hole may be constructed from  this Wightman function~\cite{Lifschytz:1994}, using the method of images:
\begin{align}
W_{\rm BTZ}(x,x') &= \sum_{n=-\infty}^\infty  \, W_{\rm AdS_3}(x, \Gamma^n x'), \label{BTZWightman}
\end{align}
where $W_{\rm AdS_3}(x, x')$ is the vacuum Wightman function associated with a massless conformally coupled scalar field in  AdS$_3$ (discussed above) and $\Gamma x'$ denotes the action of the identification  on the spacetime point $x'$.

Explicitly, for two spacetime points $x$ and $x'$ outside the black hole horizon,
\begin{align}
W_{\rm {BTZ}}(x,    x') =  \frac{1}{4 \pi \sqrt{2} \ell } \sum_{n=-\infty}^\infty\left[ \frac{1}{\sqrt{\sigma_n}} -  \frac{\zeta}{\sqrt{\sigma_n + 2}} \right],
\label{BTZWightman}
\end{align}
where
\begin{align}
\sigma_n &\ce  \frac{r r'}{\rh^2} \cosh\! \left[\frac{\rh}{\ell} ( \Delta \phi - 2 \pi n) \right] -1 - \frac{\sqrt{ (r^2 - \rh^2)(r'^2 - \rh^2)}}{\rh^2}  \cosh \!\left[\frac{\rh}{\ell^2} \Delta t \right],   \label{BTZsigma1}
\end{align}
where $\Delta \phi \ce \phi-\phi'$ and $\Delta t \ce t-t'$.

\section{Response function for AdS Rindler}

Let us begin by deriving the response function presented in Eq.~\eqref{RindResponse}. Our starting point is the definition of the response function
\be
{\cal F} \ce \frac{1}{\sigma} \int d\tau  d \tau' \, \chi_D(\tau) \chi_D(\tau')  e^{-i \Omega \left(\tau-\tau'\right)} W\!\left(x(\tau),x(\tau')\right) \,
\ee
The domain of integration is $\tau, \tau' \in \mathbb{R}^2$. We define new integration variables $s = \tau - \tau'$ and $u = \tau$. The Jacobian of this transformation is unity, allowing us to write
\be
{\cal F} = \frac{1}{\sigma} \int_{-\infty}^\infty du \chi(u) \int_{-\infty}^\infty ds \chi(u-s) e^{-i \Omega s} W(u, u-s) \, .
\ee
%From this expression it is possible using simple manipulations~\cite{Schlicht:2003iy} to express the response function in the commonly used form
%\be
%{\cal F} = \frac{2}{\sigma} {\rm Re} \int_{-\infty}^\infty du \chi(u) \int_{0}^\infty ds \chi(u-s) e^{-i \Omega s} W(u, u-s) \, .
%\ee
%However, for the present purposes it is convenient to maintain the full domain of integration for $s$.

The explicit coordinate mapping between the embedding space and the AdS-Rindler metric is given by
\begin{eqnarray}
T_1 &=& \ell \sqrt{\frac{r^2}{\ell^2}} \cosh \Phi \, ,
    \quad
X_1 =   \ell \sqrt{\frac{r^2}{\ell^2}} \sinh \Phi \, ,
    \nn\\
T_2 &=& \ell \sqrt{\frac{r^2}{\ell^2}-1} \sinh \frac{t}{\ell} \, ,
    \quad
X_2 = \ell \sqrt{\frac{r^2}{\ell^2}-1} \cosh \frac{t}{\ell} \, .
\end{eqnarray}
To write an explicit expression for the Wightman function, we note that the squared geodesic distance pulled back to the trajectory of the static detector given in Eq.~\eqref{eqn:RindAcc} is
\be
\sigma(x,x') = -2\left(\frac{R_D^2}{\ell^2} - 1 \right)  \sinh^2\frac{(t - t')}{2\ell}  =  -2 \left(\frac{R_D^2}{\ell^2} - 1 \right)  \sinh^2\frac{(\tau - \tau')}{2 \sqrt{f_{\rm AdS}(R_D)} \ell} \, ,
\ee
where $f_{\rm AdS}(R_D) \ce  \frac{R_D^2}{\ell^2}-1$. Now substituting in the Gaussian switching function, and noting that the Wightman function depends only on $s$, it is possible to immediately perform the $u$ integration yielding
\be
{\cal F} = \sqrt{\pi} \int_{-\infty}^\infty ds e^{-s^2/(4 \sigma^2)} e^{-i \Omega s} W(s) \, .
\ee
It is convenient to introduce a new coordinate $z \ce s/(2 \sqrt{f_{\rm AdS}(R_D)} \ell)$, in terms of which the remaining integral reads
\be
{\cal F} = 2 \sqrt{\pi f_{\rm AdS}(R_D)} \ell \int_{-\infty}^\infty dz e^{-f_{\rm AdS}(R_D) \ell^2 z^2/ \sigma^2} e^{-2 i \Omega \sqrt{f_{\rm AdS}(R_D)} \ell z} W(z) \, .
\ee 
Substituting in the explicit expression for the Wightman function, we obtain the result
\be
{\cal F} =   \frac{1}{4\sqrt{\pi}} \int_{-\infty}^\infty dz e^{-f_{\rm AdS}(R) \ell^2 z^2/ \sigma^2} e^{-2 i \Omega \sqrt{f_{\rm AdS}(R_D)} \ell z} \left[\frac{1}{\sqrt{-\sinh^2 z}} - \frac{\zeta}{\sqrt{-\sinh^2 z + \ell^2/(R_D^2 - \ell^2)}} \right] \, .
\ee
The second term is integrable, while the first term here has a pole at $z = 0$. To write that integral in a form amenable to numerical integration, we employ the following form of the Sokhotski formula:
\be
\frac{1}{\sinh (x-i\epsilon)} = i \pi \delta (x) + {\rm PV} \frac{1}{\sinh x} \, ,
\ee
where $\delta(x)$ is the Dirac delta distribution and ${\rm PV}$ denotes the Cauchy principal value integral. 
%\zjl{Should Sokhotskii formula have a factor of $-i\epsilon$ in order to avoid some ambiguity?  e.g.,   for  generalized function $1/ x^2$, we have $1/(x-i\epsilon)^2=1/x^2-i \pi\delta^{(1)}(x)$. I have not seen such a statement ${\rm PV}\frac{1}{x^2}$.}We therefore obtain
\be
{\cal F} =  \frac{\sqrt{\pi}}{4} - \frac{i}{4 \sqrt{\pi}} {\rm PV} \int_{-\infty}^\infty dz \frac{e^{-f_{\rm AdS}(R_D) \ell^2 z^2/ \sigma^2} e^{-2 i \Omega \sqrt{f_{\rm AdS}(R_D)} \ell z}}{\sinh z} - \frac{\zeta}{4 \sqrt{2 \pi}}  \int_{-\infty}^\infty dz \frac{e^{- f_{\rm AdS}(R_D) \ell^2 z^2/(4 \sigma^2)} e^{-i \Omega \sqrt{f_{\rm AdS}(R_D)} \ell z}}{\sqrt{1 + 2\ell^2 (R_D^2-\ell^2)^{-1} - \cosh z }} \, .
\ee
Replacing everywhere in this result the position $R_D$ in terms of the KMS temperature $T_{\rm KMS} = (2\pi \ell \sqrt{f_{\rm AdS}(R_D)})^{-1}$, and converting the domain of the second integral to $\mathbb{R}^+$, we obtain the response function quoted in the main text,
\be
{\cal F}_{\rm AdS-R} =  \frac{\sqrt{\pi}}{4} - \frac{i}{4 \sqrt{\pi}} {\rm PV} \int_{-\infty}^\infty dz \frac{e^{-z^2/(2 \pi T_{\rm KMS}\sigma)^2} e^{- i \Omega  z/ (\pi T_{\rm KMS})}}{\sinh z} - \frac{\zeta}{2 \sqrt{2 \pi}}  {\rm Re} \int_{0}^\infty dz \frac{e^{- z^2/(4 \pi T_{\rm KMS}\sigma)^2} e^{-i \Omega z/ (2 \pi T_{\rm KMS})}}{\sqrt{1 + 8 \pi^2 \ell^2 T_{\rm KMS}^2 - \cosh z }} \, .
\ee

\section{Response function for BTZ spacetime}
\label{BTZAppendix}

Next we wish to derive the response function for a detector at fixed $r=R_D$ and $\phi = 0$ in the BTZ spacetime. The starting point is again the definition of the response function, and the procedure is identical to the AdS-Rindler case above until we explicitly refer to the Wightman function. The BTZ metric can be obtained from the embedding picture via the following coordinate choices:
\begin{align}
T_1 &= \ell \sqrt{\frac{r^2}{M \ell^2}} \cosh (\sqrt{M} \phi) \, , \quad
X_1 =   \ell \sqrt{\frac{r^2}{M \ell^2}} \sinh (\sqrt{M} \phi) \, ,
\nn\\
T_2 &= \ell \sqrt{\frac{r^2}{M \ell^2}-1} \sinh \frac{\sqrt{M} t}{\ell} \, ,    \quad
X_2 = \ell \sqrt{\frac{r^2}{M \ell^2}-1} \cosh \frac{\sqrt{M} t}{\ell} \, .
\end{align}
Pulled back to the trajectory of the static detector, Eq.~\eqref{BTZsigma1} yields
\begin{align}
\sigma_n &=  \frac{R_D^2}{\rh^2} \cosh\! \left[\frac{\rh}{\ell} ( 2 \pi n) \right] -1 - \frac{ (R_D^2 - \rh^2)}{\rh^2}  \cosh \!\left[\frac{\rh}{\ell^2} \Delta t \right],
\nn\\
&= \frac{R_D^2}{\rh^2} \cosh\! \left[\frac{\rh}{\ell} ( 2 \pi n) \right] -1 - \frac{ (R_D^2 - \rh^2)}{\rh^2}  \cosh \!\left[\frac{\rh}{\sqrt{f(R_D)} \ell^2} \Delta \tau \right] \, .
\label{BTZsigma}
\end{align}
Using Eq.~\eqref{BTZsigma} we may evaluate the image sum defining the Wightman function in Eq.~\eqref{BTZWightman}. Let us treat the $n = 0$ term separately from the $n \neq 0$ terms, as it is only the term that contains a pole requiring more careful treatment. We perform essentially identical manipulations as those performed above in the AdS-Rindler case, defining instead $z \ce \rh s/(2 \sqrt{f(R_D)} \ell^2)$. Expressing the final result in terms of $T_{\rm KMS} = \rh/(2 \pi \ell^2 \sqrt{f(R_D)})$, the $n=0$ contribution to the BTZ response function is given by
\be
{\cal F}_{\rm BTZ}^{(n=0)} = \frac{\sqrt{\pi}}{4}  - \frac{i}{4 \sqrt{\pi}} {\rm PV} \int_{-\infty}^\infty dz \frac{e^{-z^2/(2 \pi T_{\rm KMS}\sigma)^2} e^{- i \Omega  z/ (\pi T_{\rm KMS})}}{\sinh z} - \frac{\zeta}{2 \sqrt{2 \pi}}  {\rm Re} \int_{0}^\infty dz \frac{e^{- z^2/(4 \pi T_{\rm KMS}\sigma)^2} e^{-i \Omega z/ (2 \pi T_{\rm KMS})}}{\sqrt{1 + 8 \pi^2 \ell^2 T_{\rm KMS}^2 - \cosh z }} \, .
\ee
For equal values of $T_{\rm KMS}$ the $n=0$ contribution to the BTZ response function is identical to the AdS-Rindler response function. This is essentially the statement that the dominant contribution to the BTZ response function arises because the black hole horizon `looks like' an acceleration horizon.

The integrals for the higher-order in $n$ terms present no obstructions to direct numerical integration. Massaging these expressions into a simpler form, we obtain the following full expression for the BTZ response function:
\begin{align}
{\cal F}_{\rm BTZ} =& {\cal F}_{\rm AdS-R}+\frac{1}{\sqrt{2 \pi}} \sum_{n=1}^{\infty} \Bigg\{\int_{0}^{\infty} dz\ {\rm{Re}}\Bigg[ \frac{\exp\big(-z^2/(4 \pi \sigma T_{\rm KMS})^2\big)\exp\big(-i \Omega z/(2 \pi T_{\rm KMS}) \big)}{\sqrt{\cosh\big(\alpha_{n}^-\big)-\cosh(z)}}\Bigg]
\nn\\
&- \zeta\int_{0}^{\infty} dz\ {\rm{Re}}\Bigg[\frac{\exp\big(-z^2/(4 \pi \sigma T_{\rm KMS})^2\big)\exp\big(-i\Omega z/(2 \pi T_{\rm KMS})\big)}{\sqrt{\cosh\big(\alpha_{n}^+\big)-\cosh(z)}}\Bigg]\Bigg\},
\end{align}
where
\begin{align}
\cosh \alpha^\mp_n &=  \left(1 + 4 \pi^2 \ell^2 T_{\rm KMS}^2 \right)\cosh (2 \pi n \sqrt{M}) \mp 4 \pi^2 \ell^2 T_{\rm KMS}^2 \, .
\end{align}
This expression for the response functions makes very clear the fact that the mass dependence of the response appears only in the higher-order in $n$ terms. It is these terms that distinguish the BTZ black hole from a `mere' acceleration horizon.

%\begin{align*}
%  \kappa_D &:= \frac{\lambda^2\sigma^2}{2}, &
%  K_D &:= \frac{\lambda^2\sigma}{2\sqrt{2\pi}}, \\
%  T_D &:= \frac{r_h}{2\pi \ell\sqrt{R_D^2-r_h^2}}, &
%  a_D &:= \frac{\gamma_D^2 \ell^4}{4\sigma^2r_h^2},\\
%  \beta_D &:= \frac{\gamma_D\Omega\ell^2}{r_h}, &
%  \alpha_{D,n}^{\mp} &:= \arccosh\left[\frac{r_h^2}{R_D^2-r_h^2}\left(\frac{R_D^2}{r_h^2}\cosh\left(\frac{r_h}{\ell}2\pi n \right)\mp1\right)\right].
%\end{align*}

\section{Infinite interaction response functions}

In the limit of infinite interaction times the response functions for both AdS-Rindler and BTZ can be obtained directly in terms of special functions. The expressions provided in the main text follow immediately  from the above expressions for the response functions, along with the integral definitions of the associated Legendre functions (see Sec. 8.715 of Ref.~\cite{gradshteyn2007})
\begin{align}
P^\mu_\nu (\cosh\alpha) &= \frac{\sqrt{2} \sinh^\mu \alpha}{\sqrt{\pi} \Gamma(\tfrac{1}{2} - \mu)} \int_0^\alpha \frac{\cosh \left[ \left(\nu + \tfrac{1}{2} \right) t \right] dt}{\left(\cosh \alpha - \cosh t \right)^{\mu + 1/2}}  \qquad \mbox{ for } \alpha>0 \mbox{ and }  \Re\left[\mu\right] < \frac{1}{2},
\nn\\
Q^\mu_\nu (\cosh \alpha) &= \sqrt{\frac{\pi}{2}} \frac{e^{\mu \pi i} \sinh^\mu \alpha }{\Gamma(\tfrac{1}{2} - \mu )} \int_\alpha^\infty \frac{e^{-(\nu + \tfrac{1}{2}) t} dt }{\left( \cosh t - \cosh\alpha \right)^{\mu + 1/2 }}  \qquad \mbox{ for } \alpha>0, \   \Re\left[\mu\right] < \frac{1}{2}, \mbox{ and } \Re\left[\nu + \mu\right],
\end{align}
and the following two identities for  associated Legendre functions of the first kind 
\begin{align}
Q_{-\nu - 1} &= Q_\nu - \pi  \cot (\pi \nu)  P_\nu \qquad \mbox{ for } \sin(\pi\nu)\ne0 \nn\\
{\rm Im} \left[Q_{-\frac{1}{2} + i \lambda} \right] &= - {\rm Im} \left[Q_{-\frac{1}{2} - i \lambda} \right] \, .
\end{align}
We note that when using the built-in associated Legendre functions in Mathematica, the branch cut structure should be taken to be type 3.

\twocolumngrid

\end{document}